# Antibacterial properties of nonwoven wound dressings coated with Manuka honey or methylglyoxal

**Sophie E L Bulman,\* Giuseppe Tronci, Parikshit Goswami, Chris Carr and Stephen J Russell**

Clothworkers' Centre for Textile Materials Innovation for Healthcare (CCTMIH), Clothworkers Central Building, School of Design, University of Leeds, Leeds, West Yorkshire, LS2 9JT; s.j.russell@leeds.ac.uk

\* selb_8@hotmail.com ; Tel.: +44 7837 913 176

**Abstract:** Manuka honey (MH) is used as an antibacterial agent in bioactive wound dressings via direct impregnation onto a suitable substrate. MH provides unique antibacterial activity when compared with conventional honeys, owing partly to one of its constituents, methylglyoxal (MGO). Aiming to investigate an antibiotic-free antimicrobial strategy, we studied the antibacterial activity of both MH and MGO (at equivalent MGO concentrations) when applied as a physical coating to a nonwoven fabric wound dressing. When physically coated on to a cellulosic hydroentangled nonwoven fabric, it was found that concentrations of 0.0054 mg cm$^{-2}$ of MGO in the form of MH and MGO was sufficient to achieve 100 CFU% bacteria reduction against gram-positive *Staphylococcus aureus* and gram-negative *Klebsiella pneumoniae*, based on BS EN ISO 20743:2007. A 3- to 20- fold increase in MGO concentration (0.0170 – 0.1 mg cm$^{-2}$) was required to facilitate a good antibacterial effect (based on BS EN ISO 20645:2004) in terms of zone of inhibition and lack of growth under the sample. The minimum inhibitory concentration (MIC) and minimum bactericidal concentration (MBC) was also assessed for MGO in liquid form against three prevalent wound and healthcare-associated pathogens, i.e. *Staphylococcus aureus*, gram-negative *Pseudomonas aeruginosa* and gram-positive *Enterococcus faecalis*. Other than the case of MGO-containing fabrics, solutions with much higher MGO concentrations (128 mg L$^{-1}$ – 1024 mg L$^{-1}$) were required to provide either a bacteriostatic or bactericidal effect. The results presented in this study therefore demonstrate the relevance of MGO-based coating as an environment-friendly strategy for the design of functional dressings with antibiotic-free antimicrobial chemistries.

## 1. Introduction

With increasing bacterial resistance to antibiotics [1-3] and concern to find alternative treatments, [3-4] the antibacterial activity of MH is of growing interest and well documented. MH inhibits the growth of clinically-relevant pathogens and biofilms found in wounds, including gram-positive strains such as Methicillin-resistant *Staphylococcus aureus* (*MRSA*) [5, 6], *and Streptococcus pyogenes* [7], and gram-negative strains including *Esherichia.coli* [8], *Proteus mirabilis* and *Enterobacter cloacae* [9], and *P.aeruginosa* [10, 11]. Gastrointestinal pathogens [12] and oral infections [13] have also shown susceptibility to MH. The effect of MH on cells required for healing, including fibroblasts and keratinocytes also suggests that MH is considered a safe compound for topical treatment [14-15]. Methylglyoxal (MGO) is a keto-aldehyde, found as a yellow liquid and present in a variety of beverages and foods including wine, beer [16], bread [17], soya, coffee, teas [18] and notably, MH [19]. Mavric et al [20] reported that MGO is responsible for the heightened and unique non-peroxide antibacterial activity associated with MH, and the minimum inhibitory concentrations (MIC) of MGO in the form of both MH and isolated synthetic compound required to have an antibacterial effect have been established. For MGO, a MIC of 1.1 mM is required to induce an antibacterial effect, whilst a range of MIC values has been observed in the case of MH, in light of inherent variations in MGO content. For example, five MHs with MGO concentrations ranging between 347 to 761 ± 25 mg kg$^{-1}$ were shown to exhibit an antibacterial effect when the MH was diluted to 15 to 30% (w/v). These resulting MGO concentrations correspond to MIC values between 1.1 mM and 1.8 mM, and therefore compare with the 1.1 mM MIC value associated with synthetic MGO [20]. The antibacterial activity of MGO in the form of solution [20, 21], hydrogel [21], polymer-based formulation [22], and poly (vinyl alcohol) fibres [23] has also been studied. With respect to MH, MGO has attracted attention because of its ability to act as a lone compound at defined concentration for the inhibition of bacterial growth, as well as its carcinostatic properties [24-28] and anti-proliferative effects on leukaemia cells [29, 30].

The antibacterial effects of MH and MGO in the form of nonwoven fabric coating have not previously been compared in terms of concentration per unit area. This is particularly important when designing dressings, where the required concentration of the active compound per unit area should be known. Nonwovens in this context relate to textile materials produced by drylaid methods, which are most commonly employed to manufacture wound dressings [31, 20].

Therefore it is of interest to understand the degree to which MGO exhibits an equivalent antibacterial effect to MH, aiming to identify a synthetically-defined alternative to MH towards the design of antibacterial dressings. Consequently, the aim of this work was firstly to compare and evaluate the antibacterial efficiency of both MH and MGO when applied as a coating to a nonwoven fabric at equivalent MGO concentrations. Secondly, we wanted to determine the minimum inhibitory concentration (MIC) and minimum bactericidal concentration (MBC) of MGO (isolated synthetic compound) in liquid form against three of the most common wound pathogens including *Staphylococcus aureus* (*S.aureus*), *Peudomonas aeruginosa* (*P. aeruginosa*) and *Enterococcus faecalis* (*E. faecalis*) [33].

## 2. Results and Discussion

Sample nomenclature used in this study is as follows: samples are coded as 'MH1, MH2, MGO 1 and MGO2', whereby 'MH or MGO' identifies the additive that the nonwoven samples were prepares with, ie. MH or MGO; and '1 or 2' describes the concentration formulation, as described in Table 1. 'NW' and 'WP' indicate the coating-free nonwoven and woven polyester control samples, respectively. Table 1 provides an overview of the different sample additive formulations.

**Table 1:** MGO concentration in either the coating solutions ($C_s$) or resulting coated nonwoven fabrics ($C_f$).

| Sample ID | $C_s$ (mg g$^{-1}$) | $C_f$ (mg cm$^{-2}$) |
|---|---|---|
| MH1 | 0.11 | 0.0057 |
| MH2 | 0.33 | 0.0169 |
| MGO1 | 0.11 | 0.0054 |
| MGO2 | 0.33 | 0.0170 |

## 2.1. Antibacterial performance of the nonwoven coated samples

### 2.1.1 Antibacterial performance of the nonwoven coated samples Using BS EN ISO 20743:2007

With reference to BS EN ISO 20743:2007 [34], the results in Tables 2 and 3 report the average reduction of bacteria in colony forming units (CFU), for either *S. aureus* or *K. Pneumoniae*, respectively. For the MH coated and synthetic MGO coated nonwovens, 100 CFU% reduction in bacteria was achieved for all samples where the calculated concentration of MGO ranged from 0.0054 mg cm$^{-2}$ to 0.0170 mg cm$^{-2}$, regardless of the strain tested.. Interestingly, an average reduction of 97 CFU% in bacteria was still reported for the nonwoven control against *S.aureus* (Table 2), whilst considerably high growth (-252 CFU%) was reported when the same sample was challenged with *K. pneumonia* (Table 3). This latter effect was still observed in the case of woven polyester control following contact with either *S. aureus* or *K. Pneumoniae* (-22438 CFU% and -5635 CFU%).

Previously, it has been reported that TENCEL® or lyocell fibres are able to reduce the growth of *S.aureus* considerably when compared with synthetic fibres such as polypropylene, polyester and polyacrylate [35]. The previous study showed that the synthetic samples exhibited 100 to 1000 times higher bacteria growth when compared with lyocell. It is conceivable that the reduced growth of bacteria observed with lyocell fibres is associated with the behaviour of the fibres in water. In the case of the synthetic fibres, there is limited penetration of water into the fibres and interactions are mainly at the surface which is fully accessible to bacterial organisms. However, because of the nano-fibrillar structure of lyocell fibres, water can be absorbed into the micro capillaries inside the fibre, such that there is a reduced life sustaining environment for the bacteria to thrive [35]. It was reported that approximately 1,333,000 nanofibrils with a diameter of 10 nm are apparent in a single TENCEL fibre, thus contributing to the highly absorbent characteristic nature of the fibre [36]. This behaviour is therefore a likely explanation as to why a reduced bacterial count (97 CFU%) was observed for the NW control in the case of *S.aureus* in the present study. Following these considerations, the thinner peptidoglycan and additional lipopolysaccharide layer present in gram-negative *K. pneumoniae* compared to gram-positive *S. aureus* [37] are likely to provide the former bacteria with increased adaptability on hydrated fibres in the experimental conditions investigated, explaining why *K. pneumoniae* growth, rather than reduction, was observed in contact with the nonwoven, similarly to the polyester, control (Table 3).

**Table 2:** Average reduction in colony forming units (CFU) for *S. aureus*. "-" indicates bacteria growth.

| Sample ID* | Average CFU immediately after inoculation | Average CFU after 18 h in incubation | Average percentage reduction (CFU%) |
|---|---|---|---|
| NW | $2.64 \times 10^4$ | $8.60 \times 10^2$ | 97 |
| WP | $1.30 \times 10^5$ | $2.93 \times 10^7$ | -22438 |
| MH1 | $3.15 \times 10^4$ | 0 | 100 |
| MH2 | $3.90 \times 10^4$ | 0 | 100 |
| MGO1 | $3.20 \times 10^4$ | 0 | 100 |
| MGO2 | $3.05 \times 10^4$ | 0 | 100 |

**Table 3:** Average reduction in colony forming units (CFU) for *K. Pneumoniae*. "-" indicates bacteria growth.

| Sample ID* | Average CFU immediately after inoculation | Average CFU after 18 h in incubation | Average percentage reduction (CFU%) |
|---|---|---|---|
| NW | $8.53 \times 10^4$ | $2.40 \times 10^5$ | -252 |
| WP | $6.80 \times 10^4$ | $3.90 \times 10^6$ | -5635 |
| MH1 | $7.07 \times 10^4$ | 0 | 100 |
| MH2 | $8.60 \times 10^4$ | 0 | 100 |
| MGO1 | $7.20 \times 10^4$ | 0 | 100 |
| MGO2 | $9.93 \times 10^4$ | 0 | 100 |

**2.1.2 Antibacterial performance of the coated nonwoven samples using BS EN ISO 20645:2004**

Tables 4 and 5 summarise the results for the NW control and MH and MGO coated nonwoven samples against *E. coli* and *S. aureus* respectively in accordance with BS EN ISO 20645:2004 [38]. Figure 1 illustrates the influence of the NW control samples on the growth of bacteria. Figures 2 and 3 exemplify the effects that MH and MGO coated samples have on the bacterial growth at varying MGO concentrations.

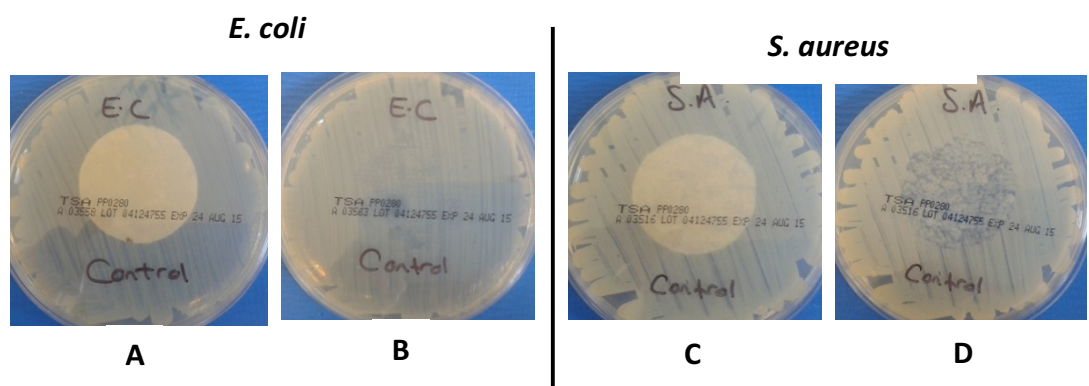

**Figure 1:** Effect of control samples on the growth of E. coli during (A and C) and following (B and D) incubation; A = no inhibition zone, B = heavy growth under sample and S. aureus; C = no inhibition zone and D = heavy growth under sample. Note: all samples were 3 cm in diameter.

**Table 4:** Effect of MGO concentration on the growth of *E. coli* when applied as a physical coating onto nonwoven samples.

| MGO concentration (mg cm$^{-2}$) | MH coatings | | | MGO coatings | | |
|---|---|---|---|---|---|---|
| | Inhibition zone (mm) | Growth under sample | Assessment | Inhibition zone (mm) | Growth under sample | Assessment |
| NW | 0 | Heavy | Insufficient | 0 | Complete | Insufficient |
| 0.0054 | 0 | Heavy | Insufficient | 0 | Moderate | Insufficient |
| 0.0170 | 0 | Slight | Limited efficiency | 0 | No growth | Good effect |
| 0.10 | 0 – 1 | No growth | Good effect | 0 | No growth | Good effect |
| 0.15 | 0 - 1 | No growth | Good effect | 0 | No growth | Good effect |
| 0.20 | >1 | No growth | Good effect | 0 | No growth | Good effect |
| 0.40 | n/a*$^1$ | n/a | n/a | 0 - 1 | No growth | Good effect |
| 0.80 | n/a*$^1$ | n/a | n/a | >1 | No growth | Good effect |
| 1.20 | n/a*$^1$ | n/a | n/a | >1 | No growth | Good effect |

*$^1$: Owing to the viscosity of the Manuka honey, it was not possible to prepare samples at concentrations above 0.2 mg cm$^{-2}$.

**Table 5: Effect of MGO concentration on the growth of *S. aureus* when applied as a physical coating onto nonwoven samples.**

| MGO concentration (mg cm$^{-2}$) | MH coatings | | | MGO coatings | | |
|---|---|---|---|---|---|---|
| | Inhibition zone (mm) | Growth under sample | Assessment | Inhibition zone (mm) | Growth under sample | Assessment |
| NW | 0 | Heavy | Insufficient | 0 | Heavy | Insufficient |
| 0.0054 | 0 | Heavy | Insufficient | 0 | Moderate | Insufficient |
| 0.0170 | 0 | Heavy | Insufficient | 0 | Slight | Limit of efficiency |
| 0.10 | 0 – 1 | No growth | Good effect | 0 | No growth | Good effect |
| 0.15 | >1 | No growth | Good effect | 0 | No growth | Good effect |
| 0.20 | >1 | No growth | Good effect | 0 | No growth | Good effect |
| 0.40 | n/a*1 | n/a | n/a | 0 | No growth | Good effect |
| 0.80 | n/a*1 | n/a | n/a | >1 | No growth | Good effect |
| 1.20 | n/a*1 | n/a | n/a | >1 | No growth | Good effect |

*1: Owing to the viscosity of the Manuka honey, it was not possible to prepare samples at concentrations above 0.2 mg cm$^{-2}$.

As shown in Fig. 1A and 1C, no zone of inhibition was apparent with the control samples when tested against both gram-negative *E.coli* and gram-positive *S.aureus.* Upon removal of the control samples from the surface of the agar, the contact zone between the sample and the agar presented heavy bacterial growth (Fig. 1B & 1D). This confirms that the control samples did not exhibit any antibacterial activity. Whilst these observations appear to be in contrast with the results provided in Table 2, it is important to note that in this case, samples were directly tested in contact with inoculated agar gels, in the absence of simulated wound exudate solution (in contrast to the case of the assay results provided in Table 2). Here, the bacteria detrimental fibre-induced water uptake effect was largely marginal, so that high growth of *S.aureus* was consequently still observed following application of the nonwoven control sample. The antibacterial effect of the MH and MGO coatings having equivalent MGO concentrations between 0.0054 mg cm$^{-2}$ and 0.0170 mg cm$^{-2}$ showed no zone of inhibition for *E.coli* and *S.aureus.* Upon removal of the MH coated samples from the agar, heavy growth was apparent at an MGO concentration of 0.0054 mg cm$^{-2}$ for both *E.coli* and *S.aureus* (Fig. 2A displays an example of heavy growth). Moderate growth was achieved for both *E.coli* and *S.aureus,* upon removal of the MGO coated nonwovens at equivalent concentrations (an example of moderate growth is shown in Fig. 2B). At an MGO concentration of 0.0170 mg cm$^{-2}$, moderate and heavy growth was observed for *E.coli* and *S.aureus* respectively for the MH coated samples. However, for the MGO coatings with an equivalent MGO concentration of 0.0170 mg cm$^{-2}$, no growth and slight growth was evident against *E.coli* and *S.aureus*, respectively (examples of slight growth and no growth are shown in Fig. 2C and Fig. 2D). These initial evaluations at a concentration of 0.0054 mg cm$^{-2}$ suggest an insufficient antibacterial effect was achieved for both MH and MGO coatings. At a concentration of 0.0170 mg cm$^{-2}$, limited efficacy was observed for the MH coatings. However, for the MGO coatings with an MGO concentration of 0.0170 mg cm$^{-2}$, the antibacterial effect was shown to improve slightly and a good antibacterial effect and a limit of efficiency was achieved for both *E.coli* and *S.aureus* respectively.

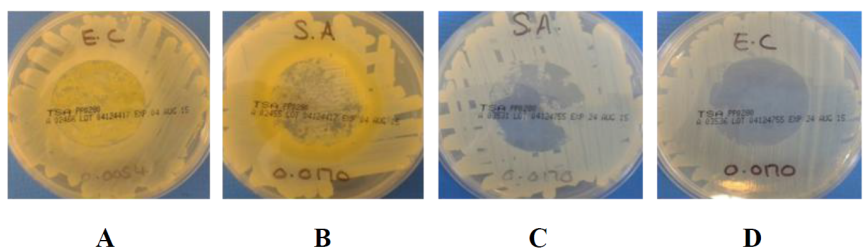

      **A**            **B**          **C**          **D**

**Figure 2: Examples of bacteria growth under either MH or MGO coated nonwoven samples. A = heavy growth of *E. coli*, B = medium growth of *S. Aureus*, C = slight growth of *S. aureus* and D = no growth of *E. coli*. Note: all coated samples were 3 cm in diameter.**

It is important to note that where no growth or inhibition zone was apparent, a good antibacterial effect may still be observed. This may be linked to the diffusion rate of the active compound from the fabric [38] to the agar plate and the affinity of the fibres for moisture. Thus, it is likely that, within the time frames investigated in this study, herein the hygroscopic, crystalline nano-fibrils of the TENCEL® fibres [35] retain the MGO and honey coating, thereby limiting the diffusion of MGO into the agar at these MGO concentrations. This situation may well be expected in this case, given that no additional simulated wound exudate solution was applied. The minimal swelling of the fibres expected following contact with the agar plate may well be directly related to a decreased MGO diffusion. This hypothesis is supported when comparing data obtained in exudate-free conditions with the ones presented in Tables 2 and 3, where complete bacteria killing was observed with the same MGO concentrations following addition of simulated wound exudate solution. As the MGO concentration increases between 0.1 mg cm$^{-2}$ and 1.2 mg cm$^{-2}$, a good antibacterial effect is observed with both MH and MGO in all cases (Table 4 and 5). For the MH coated samples, mean zones of inhibition of 0-1mm were apparent against both *E.coli* and *S.aureus* at concentrations between 0.1 mg cm$^{-2}$ and 0.2 mg cm$^{-2}$. Fig. 3B displays an example of an inhibition zone from 0-1mm. As the concentration of MGO doubled to 0.2 mg cm$^{-2}$, the mean zone of inhibition for *E.coli and S.aureus* increased to achieve a mean zone of >1mm. An example of a mean zone of >1mm can be seen in Fig. 3C.

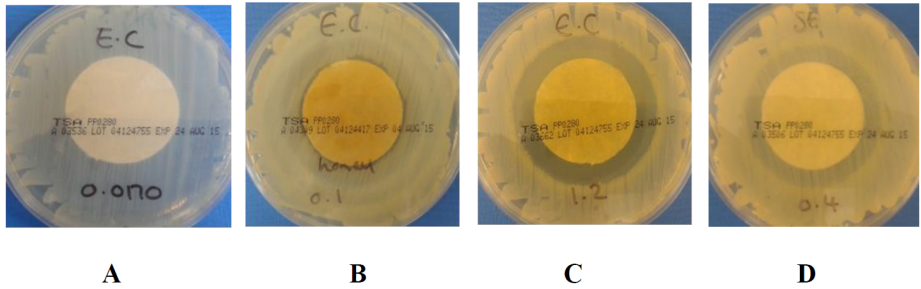

**Figure 3:** Examples of zones of inhibition formed around both MH and MGO coated nonwoven samples. A = no zone (*E. coli*), B = 0-1mm (*E. coli*), C = >1mm (*E. coli*) and D = unclear zone (*S. aureus*). Note: all coated samples were 3 cm in diameter.

Conversely, the MGO coatings did not show a clear zone of inhibition until a concentration of 0.4 mg cm$^{-2}$ was reached for *E. coli* and 0.8 mg cm$^{-2}$ for *S. aureus*. Below these concentrations, no evidence of bacterial growth was observed upon removal of the samples, resulting in a good antibacterial effect. However, a partial zone of inhibition was formed around the samples, as presented in Fig. 3D, suggesting the TENCEL® fibres still retained a proportion of the MGO. As the addition of MGO solution increased, the TENCEL® fibres uptake of, and ability to retain, the MGO was reduced. This is expected to encourage greater diffusion of MGO into the bacteria-seeded agar, resulting in a clear zone of inhibition. Fig. 4 shows FEGSEM images of the dry (Fig. 4A), MGO coated (Fig. 4B) and the MH coated (Fig. 4C&D) TENCEL® fibres. It is apparent that the MGO coated TENCEL® fibres (Fig 4B) resemble a similar appearance to the dry TENCEL® fibres as seen in Fig. 4A, confirming that the liquid phase coating has been absorbed and retained by the fibres. The MH coated TENCEL® fibres appear mainly occluded by the honey coating (Fig 4C&D), and some protruding fibres exhibit a globular surface coating due to the MH. These images provided further evidence that the MH is freely available on the surface of the TENCEL® fibres such that direct contact with the bacteria agar can be anticipated. It is also likely that during incubation at 37ºC, the MH coating will soften and allow greater diffusion into the bacteria seeded agar from the fibres. Previous studies have reported that temperature has a direct influence on the viscosity of honey [39-41], such that as temperature increases the viscosity falls due to reduced hydrodynamic forces and reduced molecular interaction [41]. Viscosity measurements of the MH obtained during this study confirmed this temperature dependency, with a decreased viscosity of 17800 cP being obtained at 37ºC ± 2ºC rising to 21800 cP at 25 ± 2ºC. This data therefore suggests that the MH coating is more likely to migrate freely into the bacteria seeded agar during testing, as a result of low viscosity at 37ºC.

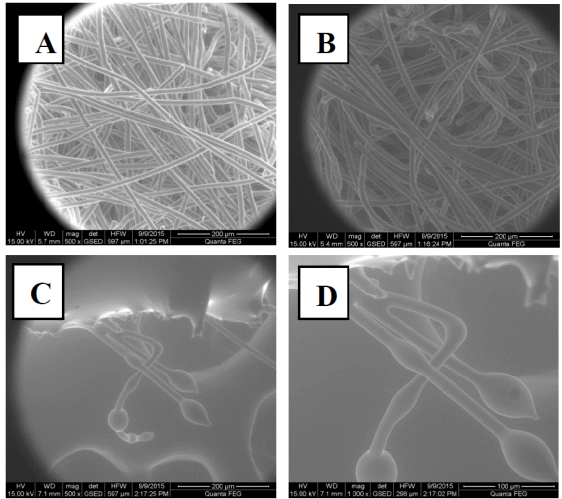

**Figure 4:** FEGSEM of dry TENCEL® fibres (A), synthetic MGO coated fibres (B) and Manuka honey coated fibres (C&D). Taken at a magnification of 500x (A, B & C) at 1000x (D). The concentration of MGO on both the MGO and MH coated samples was 0.1 mg cm². The diameter of the uncoated TENCEL fibres ranged between 10 and 15μm. MGO coated fibres ranged from 10μm to 25μm (this is due to the swelling of the TENCEL fibres after coating).

Previous investigation of the antibacterial activity of MH and MGO in a liquid form reported that higher levels of MGO alone were required to inhibit the growth of *P.aeroginosa* when compared with MH where equivalent MGO concentrations were apparent [42]. Secondly, the presence of hydrogen peroxide in the MH may contribute to the heightened antibacterial effect [42-45].

Comparing the results obtained using both antibacterial methods, the concentration of MGO required to produce an antibacterial effect was found to be slightly lower (0.0054 mg cm²) when assessed according to BS EN ISO 20743:2007, compared to concentrations between 0.0170 mg cm² and 0.1 mg cm² for *E.coli* and *S.aureus* respectively when using BS EN ISO 20645:2004. The lower MGO concentration achieved using BS EN ISO 20743:2007 may be attributed to the addition of the liquid used to simulate wound exudate (SCDLP) solution during testing [34]. This would result in the TENCEL® fibres being exposed to higher moisture content, which could encourage hydration of the fibres and facilitate extraction of the MGO. In the case of BS EN ISO 20645:2004 an insufficient moisture content is available to initiate the diffusion of the MGO from the fibres [38]. It is only when the nonwoven samples become increasingly hydrated that diffusion of MGO into the agar is promoted.

## 2.3. MIC and MBC of MGO against common wound pathogens

In addition to coatings on nonwoven fabric, Table 6 gives the results of the MIC and MBC of MGO in liquid form against gram-positive *S.aureus*, gram-negative *P.aeruginosa* and gram-positive *E.faecalis*. For *P.aeruginosa* the MIC against the ATCC strain and patient 1 strain was found to be 512 mg L$^{-1}$. Twice the concentration was required (1024 mg L$^{-1}$) to inhibit the growth of the patient 2 strain. Upon subculture of all three isolates, the MBC required to kill *P.aeruginosa* ATCC strain was doubled to 1024 mg L$^{-1}$, while the two patient strains remained at 512 mg L$^{-1}$. These concentrations were the highest among each bacterium species tested, as the MIC for *S.aureus* ranged from 126 mg L$^{-1}$ to 256 mg L$^{-1}$. While *E. faecalis* did not reach above 512 mg L$^{-1}$. It is not surprising that the relatively high concentrations required to inhibit or kill *P.aeroginosa* were found, as *P.aeruginosa* is one of the most problematic multidrug-resistant gram-negative strains that is showing increasing resistance to common antibiotics, including ciprofloxacin [46], amikacin and imipenem [47]. Previous studies have reported limited findings on the MBC of MGO against *P.aeruginosa*. One previous study reported the MBC of both MRSA and *P.aeruginosa* in a planktonic and biofilm state. In the planktonic state, the MBC for *P.aeruginosa* was found to range between 600 mg L$^{-1}$ to 1200 mg L$^{-1}$, while in the biofilm state the MBC was found to be much higher, ranging from 1800 mg L$^{-1}$ to 7600 mg L$^{-1}$ [48]. The raised MBC of MGO can be expected in a biofilm state, as the polysaccharide (Psl) in *P.aeruginosa* biofilms has been shown to provide a physical barrier against various antibiotics at the beginning stages of biofilm development [49]. In the current study, the MBC was only tested in the planktonic state and shows a slightly lower MBC between 512 mg L$^{-1}$ and 1024 mg L$^{-1}$ than in the previous study.

The lowest MIC and MBC results were found to be against gram-positive *S.aureus*, reportedly the most common bacterium species found in a chronic wound environment [31]. The ATCC strain showed the lowest MIC of 128 mg L$^{-1}$, while the two patient strains did not exceed 256 mg L$^{-1}$ for both the MIC and MBC. Previous studies have reported a range of MGO concentrations required to inhibit or kill *S.aureus*. One study reported a lower MIC 79.3 mg L$^{-1}$ (1.1mM) [20], while another reported a biocidal effect of MGO in MH, with an MGO concentration of 530 mg L$^{-1}$ against biofilms [50]. In the same study, the concentration of MGO alone (non Manuka honey sample) required to achieve a biocidal effect against biofilms was >1050 mg L$^{-1}$, which is four times higher than the concentration reported in the current study (256 mg L$^{-1}$). However the current study has only tested the effect of MGO against *S.aureus* in the planktonic state, and so a lower concentration is to be expected.

The MIC and MBC of MGO against *E.faecalis*, has, to the authors knowledge, not been previously reported. One study reported the effect of Activon MH dressing against *vancomycin-resistant Enterococcus faecalis* (VRE), stating that a 5 % (w/v) concentration was needed to initiate an antibacterial effect against the biofilm. Yet no reference was made throughout the study in relation to MGO. In the current study, equivalent MIC and MBC were shown to be effective for the ATCC strain at a concentration of 256 mg L$^{-1}$ and the two patient strains with a concentration of 512 mg L$^{-1}$. *E.faecalis* is reported to be the second most prevalent bacteria found in a chronic wound environment [33], therefore it was of importance to discover the required concentration of MGO to provide a bacteriostatic and bactericidal effect in a planktonic state.

In light of the MIC and MBC concentrations found in this study, it is conceivable that MGO concentrations covering these ranges are likely found in Manuka dressings currently applied in clinical situations. Previous studies have only addressed the concentration of Manuka honey in a commercial dressing in terms of w/v, with no specific reference to MGO content. Therefore, this study along with the work by Mavrik et al [20] and Jervis et al [50], give an indications of the appropriate MGO concentrations required to inhibit or kill a broader spectrum of bacterial strains.

**Table 6: MIC and MBC (mg L$^{-1}$) of MGO in liquid form against three common wound pathogens.**

| Test organism | MIC (mg L$^{-1}$) | MIC (mM) | MBC (mg L$^{-1}$) | MBC (mM) |
|---|---|---|---|---|
| Pseudomonas aeruginosa ATCC27853 | 512 | 7.1 | 1024 | 14.2 |
| Pseudomonas aeruginosa patient 1 | 512 | 7.1 | 512 | 7.1 |
| Pseudomonas aeruginosa patient 2 | 1024 | 14.2 | 1024 | 14.2 |
| Staphylococcus aureus ATCC29213 | 128 | 1.8 | 256 | 3.6 |
| Staphylococcus aureus patient 1 | 256 | 3.6 | 256 | 3.6 |
| Staphylococcus aureus patient 2 | 256 | 3.6 | 256 | 3.6 |
| Enterococcus faecalis ATCC21292 | 256 | 3.6 | 256 | 3.6 |
| Enterococcus faecium (VRE) patient 1 | 512 | 7.1 | 512 | 7.1 |
| Enterococcus faecium (VRE) patient 2 | 512 | 7.1 | 512 | 7.1 |

## 3. Materials and Methods

### 3.1 Materials

A 40 wt % MGO aqueous solution was purchased from Sigma Aldrich UK. Manuka honey 550+ was purchased from Wellbeing UK and TENCEL® cellulose fibres with a linear density of 1.7 dtex and length of 10 mm were obtained from Lenzing, Austria.

### 3.2 Preparations of MH and MGO coating solutions

A 20% (w/w) and a 60% (w/w) aqueous solution of MH was prepared by dissolving 100 g and 300 g of Manuka honey (MGO 550+) respectively, in distilled water and made up to 500 g. The concentration of MGO of the two solutions was calculated as 0.11 mg g$^{-1}$ and 0.33 mg g$^{-1}$ respectively. A 40 wt% MGO solution was diluted with distilled water to obtain equivalent concentrations of MGO.

### 3.3 Preparation of coated nonwoven dressings

Prior to the manufacture of the nonwoven samples, TENCEL® fibres were opened using a Shirley fibre blender. Airlaid webs with a basis weight of 120 g m$^{-2}$ were produced from 100% TENCEL® fibre using a short fibre airlaying machine in which the fibres are sifted through a static screen aided by rotating blades. The webs were mechanically bonded by hydroentanglement (water jet entanglement) using an STL Hydrolace system with a 110 - 120 μm diameter jet strip and a jet pressure of 50 bar on one side and 50 bar on the reverse. The hydroentangled webs were washed with warm water and fabric detergent to remove residual chemical finish on the fibres after hydroentangling. Using a sample liquor ratio of 1:50, the coatings were applied by immersing samples into the prepared MH and MGO coating solutions for 10 min. The samples were then passed through a pad mangle at a pressure of 10 kg cm$^{-2}$, weighed and left to air dry at room temperature. A coating-free sample was also prepared and used as a control (NW). The amount of MGO per unit area (mg cm$^{-2}$) absorbed onto the coated nonwoven samples was calculated to range between 0.0054 mg cm$^{-2}$ and 0.0170 mg cm$^{-2}$ as indicated in Table 1. Following initial antibacterial testing at these relatively low concentrations, additional coated nonwoven samples were prepared to provide six new MGO concentrations between 0.1 mg cm$^{-2}$ and 1.2 mg cm$^{-2}$. This was achieved by the addition of either MH or MGO to the pre-made coated nonwovens. A 7 cm$^2$ sample was placed in a weighing boat and weighed on a microbalance. The addition of MH or MGO to the nonwoven sample equated to the required weight needed to give the new range of MGO concentrations between 0.1 mg cm$^{-2}$ and 1.2 mg cm$^{-2}$. Prior to addition of the Manuka honey 550+, the honey was heated in an incubator at 40ºC to allow it to soften and enable a homogeneous distribution over the nonwoven sample. Note it was not possible to prepare MH samples at concentrations above 0.2 mg cm$^{-2}$ as the nonwoven samples could not retain additional material due to the high density and viscous nature of the honey.

### 3.4 Characterisation of the nonwoven coated samples

### 3.4.1 Morphology of the coated nonwoven samples

The coated nonwoven samples were inspected using an FEI Quanta 200F Field Emission Scanning Electron Microscope (FEGSEM). Prior to imaging, all samples were cut and mounted onto 25 mm aluminium stubs and sputter coated with gold in a vacuum of 0.05 torr for 4 min at 20 mA. A voltage of 15 kV and a vacuum pressure in the order of 10$^{-6}$ mbar was achieved in the chamber. Magnifications between x500 and x1000 were used in order to record morphological features of individual nonwoven coated samples.

### 3.4.2 Viscosity measurements of the MH coatings

A Brookfield LV viscometer (DV-E) was used to measure the viscosity of MH (MGO 550+) solutions. The solution viscosity of the MH was measured at a temperature of 25ºC ± 2ºC and 37 ºC ± 2ºC to assess temperature dependency. In order for the chamber and solution to reach the specific temperature required, 16.1 ml of the MH was decanted into the chamber and conditioned in an S1 500 Orbital Incubator at the required temperature for 24 h prior to testing. The spindle was also conditioned to the correct temperature. A speed of 6 r min$^{-1}$ and a spindle size of 18 were used.

### 3.5 Antibacterial evaluation of MH and MGO-coated nonwoven samples

The antibacterial activity of the MH and MGO-coated nonwoven samples was determined using two standard methods, BS EN ISO 20743:2007 (Textiles - Determination of antibacterial activity of antibacterial finished products) [34] and BS EN ISO 20645:2004 (Textile fabrics - Determination of antibacterial activity, agar diffusion plate test) [38]. The first method simulates the effect of an antibacterial dressing in contact with contaminated wound exudate [51] and was used to indicate the antibacterial effect at equivalent MGO concentrations. The second method made an assessment of the MH and MGO coated nonwovens, as well as the NW using a bacteria seeded agar plate. A brief description of each method is given below.

**3.5.1 BS EN ISO 20743:2007 (Textiles - Determination of antibacterial activity of antibacterial finished products)**

Test pieces with a mass of 0.40 g ± 0.05 g were cut into suitable sizes for testing. Six control specimens and six antibacterial specimens were prepared, based on standard protocols [34]. In this study bacteria cultures of both *S.aureus* and *K.Pneumonia* were prepared to concentrations between $1\text{-}3\cdot10^5$ per 10 ml in 1 in 20 nutrient broth. The test specimens were placed in sterile jars and inoculated with 0.2 ml of bacterial suspension on several areas of the sample, taking care to prevent contact of the suspension with the jar surface. Immediately after inoculation, 20 ml of SCDLP medium (simulated wound exudate) was added to three of the control jars and three of the antibacterial sample jars. The jars were sealed with caps and shaken in an arc of approximately 30 cm by hand for 30 sec. The number of bacteria recovered from the samples was then determined using a standard serial dilution and pour plate technique using peptone salt solution as the dilutant and enumeration agar. The remaining jars were incubated at 37ºC for 24 h. After the incubation period, the number of bacteria that could be recovered was determined using the equation in the standard.

**3.5.2 BS EN ISO 20645:2004 (Textile fabrics - Determination of antibacterial activity, agar diffusion plate test)**

A circular specimen of fabric with a diameter of 25 ± 5 mm was cut from the test sample. Two specimens of the antibacterial fabric and two control specimens without addition of antibacterial treatment were prepared, based on standard protocols [38]. The specimens were stored between 12 h to 24 h in sterilized petri dishes at room temperature. Separate agar plates were inoculated with *S.aureus* and *E.coli* bacterial species via streaking the plates with an inoculation loop from a solution containing $1\text{-}5 \times 10^8$ colony forming units per ml. The test specimen was placed onto the bacterial inoculated agar surface using a sterilized pair of tweezers until the texture of the specimen was uniformly imprinted onto the agar. The petri dishes were placed in the incubator for 24 h at 37°C ± 1°C. Immediately after this period the petri dishes were examined for bacterial growth. If any zone of inhibition was formed around the test specimens, the diameter of the zone was measured using a pair of calibrated callipers. The microbial zone of inhibition was calculated using the equation in the standard. Table 7 shows the criteria stipulated in the standard for defining the effect of an antibacterial treatment [38].

**Table 7: Stipulated criteria for defining the effect of an antibacterial treatment.**

| Inhibition Zone (mm) | Growth | Description | Assessment |
|---|---|---|---|
| >1<br>1-0<br>0 | None<br>None<br>None | inhibition zone exceeding 1 mm, no growth [b]<br>inhibition zone up to 1 mm, no growth [b]<br>no inhibition zone, no growth [c] | Good effect |
| 0 | Slight | no inhibition zone, only some restricted colonies, growth nearly totally suppressed [d] | Limit of efficacy |
| 0<br>0 | Moderate<br>Heavy | no inhibition zone, compare to the control growth reduced to half [e]<br>no inhibition zone, compare to the control no growth reduction or only slightly reduced growth | Insufficient effect |
| [a] The growth of bacteria in the nutrient medium under the specimen.<br>[b] The extent of the inhibition shall only partly be taken into account. A large inhibition zone may indicate certain reserves of active substances or a weak fixation of a product on the substrate.<br>[c] The absence of growth, even without inhibition zone, may be regarded as a good effect, as the formation of such an inhibition zone may have been prevented by a low diffusibility of the active substance.<br>[d] "As good as no growth" indicates the limits of efficacy.<br>[e] Reduced density of bacterial growth means either the number of colonies or the colony diameter. | | | |

**3.5.3 Antibacterial evaluation of MGO solutions against common wound pathogens**

Using a standard laboratory assay [52], the MIC and MBC of MGO was determined against three common wound pathogens. A 40 wt % MGO solution was diluted to concentrations between 1 mg L$^{-1}$ to 1026 mg L$^{-1}$ by diluting a working stock solution in Mueller Hinton broth. These concentrations were chosen based on similar concentrations reviewed in the current literature [53]. The dilution series was dispensed across all rows of a 96-well microtitre tray in 50μl amounts. Three of the most common bacterial isolates found in wounds including S.*aureus*, *P.aeruginosa* and *E.faecalis* were chosen for the study. Two separate strains were collected from patients following ethical approval, and an American type culture collection (ATCC) standard of each was also used. The bacteria isolates were inoculated on to fresh blood (FBA) agar plates and incubated aerobically at 37°C for 18-24h. Single colonies of each bacterial isolate were removed from the agar plates and resuspended in 5mL Mueller Hinton Broth to a 0.5 MacFarland turbidity equivalent (1·10$^8$ CFU·mL$^{-1}$). Starting with the non-MG-containing growth control, and then working from the lowest to the highest MG-containing broth, duplicate rows of the 96 well MIC plate were inoculated with 20 μL of each bacterial isolate, to yield a bacterial concentration of 2·10$^6$·well$^{-1}$. Inoculation of the MIC plates occurred within 15 minutes of inoculum preparation and these were then incubated at 37°C for 18-24h. The MIC was defined as the lowest concentration of MGO that completely inhibited the growth of the bacterial isolates, as detected by the unaided eye. In order to determine whether the growth inhibition at any particular dilution was bactericidal or bacteriostatic, triplicate 20μl aliquots were inoculated onto each of three thirds of an FBA agar plate and spread over the surface of the FBA agar third with a sterile inoculating loop. Inoculated FBA plates were then incubated at 37°C for 18-24h. The MBC was defined as the lowest concentration at which there was no visible bacterial growth upon FBA.

## 4. Conclusions

This study provides the first comparison of equivalent MGO concentrations per unit area in MH and MGO that are required to provide an antibacterial effect when applied as a physical coating to a nonwoven wound dressing fabric. The antibacterial efficiency was investigated using both BS EN ISO 20743:2007 and BS EN ISO 20645:2004 to determine if synthetic MGO provided a comparable antibacterial effect to MH. In the first instance, the bacteria inoculated samples were immersed in 20 ml of simulated wound exudate fluid and it was found that an MGO concentration of 0.0054 mg cm$^{-2}$ for both MH and MGO was sufficient to achieve 100% reduction in bacteria when tested against Gram positive *S.aureus* and Gram negative *K.pneumonia*.

Experiments using bacteria seeded agar plates, found that higher concentrations of MGO between 0.0170 mg cm$^{-2}$ and 0.1 mg cm$^{-2}$ were required to produce a good antibacterial effect against *E.coli* and *S.aureus*. In the case of BS EN ISO 20743:2007, hygroscopic TENCEL® fibres are hydrated due to the addition of 20 ml of SCDLP, which is likely to encourage MGO diffusion, as compared to BS EN ISO 20645:2004, where samples are only incubated with agar and limited moisture is available to facilitate the diffusion mechanism.

The MH coated nonwovens produced zones of inhibition at relatively low MGO concentrations of between 0.1 mg cm$^{-2}$ and 0.2 mg cm$^{-2}$, as compared with MGO-only coated nonwovens. Clear zones of inhibition were not apparent until a MGO concentration threshold was reached of 0.4 mg cm$^{-2}$ for *E.coli* and 0.8 mg cm$^{-2}$ for *S.aureus*. This difference was attributed to the incubation of the samples at 37°C during testing, where the MH coating is likely to soften promoting more rapid diffusion into the bacteria seeded agar from the fibres, as compared to the less viscous MGO coating, more of which is retained by the TENCEL® fibres. Manuka honey also contains hydrogen peroxide, which is likely to contribute to the heightened antibacterial effect, when compared with MGO.

Limited research has previously been reported regarding the MIC and MBC of MGO. In this study, the MIC and MBC against *P.aeuroginosa* was found be lower than that previously reported in the literature when in a planktonic state. The MIC and MBC for *S.aureus* was found to be between the two previously reported results of 79.3 mg L$^{-1}$ [20] and >1050 mg L$^{-1}$ [48, 50] where the latter value relates to the antimicrobial effect of MGO against biofilms. The MIC and MBC of MGO against *E.facaelis* is reported for the first time in the present study.

**Acknowledgments:** The authors also wish to thank Dr. Jane Freeman and Mr Peter Parnell for their assistance with the antimicrobial studies at the Leeds Pathology department. The authors gratefully acknowledge the financial support of the Clothworkers' Foundation and The Clothworkers' Centre for Textile Materials Innovation for Healthcare (CCTMIH).